\def\year{2021}\relax
\documentclass[letterpaper]{article} %
\usepackage{aaai21}  %
\usepackage{times}  %
\usepackage{helvet} %
\usepackage{courier}  %
\usepackage[hyphens]{url}  %
\usepackage{graphicx} %
\usepackage{natbib}  %
\usepackage{caption} %
\usepackage{amsmath}
\usepackage{amssymb}
\usepackage{amsfonts}
\usepackage{algorithmic}
\usepackage{graphicx}
\usepackage{textcomp}
\usepackage{xcolor}
\usepackage{multirow}
\usepackage[labelformat=simple]{subcaption}
\usepackage{booktabs}
\usepackage{multirow}
\usepackage[ruled,vlined]{algorithm2e}
\usepackage{pifont}

\newcommand{\V}{\mathcal{V}}
\newcommand{\E}{\mathcal{E}}
\newcommand{\G}{\mathcal{G}}
\newcommand{\cmark}{\ding{51}}
\newcommand{\xmark}{\ding{55}}

\newcommand{\eg}{e.g., }
\newcommand{\ie}{i.e., }

\newcommand{\figref}[1]{Figure~\ref{#1}}
\newcommand{\tableref}[1]{Table~\ref{#1}} 
\newcommand{\equref}[1]{Equation~\ref{#1}} 
\newcommand{\algref}[1]{Algorithm~\ref{#1}}

\newcommand{\citea}[1]{~\citeauthor{#1}}
\newcommand{\citepp}[1]{~\citep{#1}}

\title{SDGNN: Learning Node Representation for Signed Directed Networks}

\author {
    Junjie Huang, Huawei Shen$^{*}$, Liang Hou,  Xueqi Cheng\\
}
\affiliations {
    CAS Key Laboratory of Network Data Science and Technology, Institute of Computing Technology, \\Chinese Academy of Sciences,  Beijing, China\\
University of Chinese Academy of Sciences, Beijing, China\\
\{huangjunjie17s, shenhuawei, houliang17z, cxq\}@ict.ac.cn\\
}

\begin{document}

\maketitle

\begin{abstract}
  Network embedding is aimed at mapping nodes in a network into low-dimensional vector representations.
  Graph Neural Networks (GNNs) have received widespread attention and lead to state-of-the-art performance in learning node representations.
  However, most GNNs only work in unsigned networks, where only positive links exist.
  It is not trivial to transfer these models to signed directed networks, which are widely observed in the real world yet less studied.
  In this paper, we first review two fundamental sociological theories (\ie status theory and balance theory) and conduct empirical studies on real-world datasets to analyze the social mechanism in signed directed networks.
  Guided by related sociological theories, we propose a novel \underline{S}igned \underline{D}irected \underline{G}raph \underline{N}eural \underline{N}etworks model named SDGNN to learn node embeddings for signed directed networks.
  The proposed model simultaneously reconstructs link signs, link directions, and signed directed triangles.
  We validate our model's effectiveness on five real-world datasets, which are commonly used as the benchmark for signed network embeddings.
  Experiments demonstrate the proposed model outperforms existing models, including feature-based methods, network embedding methods, and several GNN methods.

\let\thefootnote\relax\footnotetext{*Corresponding Author}
\end{abstract}
\section{Introduction}\label{sec:introduction}
With the growing popularity of online social media, many interactions between people are generated and recorded on the web.
Modeling and understanding these interactions is a useful perspective on a range of social computing studies.
Most of these interactions are positive relationships, such as friend, trust, like, support, and approval.
Meanwhile, the conflicts on the web\citepp{kumar2018community} are everywhere, forming negative links that indicate hostility, distrust, hate, disagree, and disapproval.
Since the interactions are generated by people and reflect people's attitude and psychology, related researches on social psychology (\eg status theory and balance theory) shed light on the social mechanism of signed networks.
These two theories are widely used in signed network analysis\citepp{leskovec2010predicting}.
In signed network analysis, link sign prediction is to predict the sign of a given link.
\citea{leskovec2010predicting} carefully extract features based on social theories and achieve good performances.
Link sign prediction is the main downstream machine learning task for evaluating signed network embedding.
Signed network embedding learning, which aims to map nodes in signed network to low-dimensional vector representations, also relies on these sociological theories \citepp{wang2017signed,kim2018side,chen2018bridge,islam2018signet,chen2018bassi}.
For example, \citea{chen2018bridge} mathematically model ``bridge'' edges based on balance and status theory and achieve state-of-the-art performances.

Recently, Graph Neural Networks (GNNs) have been receiving more and more attention.
Modern GNNs mostly follow Message Passing Neural Networks (MPNNs)\citepp{gilmer2017neural} manner, which consist of message aggregators and update functions.
The common aggregators include mean aggregator\citepp{hamilton2017inductive}, attention aggregator\citepp{velickovic2017graph}, and max-pooling aggregator\citepp{hamilton2017inductive}.
GNNs have achieved good results in many machine learning tasks (\eg semi-supervised node classification, learning low-dimensional representations, and link prediction)\citepp{kipf2016semi,kipf2016variational,wang2016structural}.  
A number of recent researches have pivoted to learn the node embeddings using GNNs, which aim to aggregate information from neighbors for node embeddings\citepp{kipf2016variational,duran2017learning,pan2018adversarially}. 
These GNN-based network embedding methods have revolutionized the field of network embedding and achieved state-of-the-art performances. 
Specifically, Graph Auto-Encoder (GAE) uses GCNs to process node features jointly with the graph structure to produce a set of hidden representations \citepp{kipf2016variational}. 
It uses a scoring function to reconstruct the adjacency matrix of the graph from hidden representations.
GAE is designed for the unsigned networks and can't directly be applied in the signed networks.

For signed network, SGCN generalizes GCN to signed networks and designs a new information aggregator based on balance theory\citepp{derr2018signed}.
SNEA leverages the self-attention mechanism to enhance the performance of signed network embeddings\citepp{li2020learning}.
Although signed GNNs discussed above are proposed to model signed networks with balance theory, they don't take direction (\ie status theory) into consideration, which is important for signed graph modeling\citepp{cui2020semi}.
In this paper, we try to model signed directed networks with a new \underline{S}igned \underline{D}irected \underline{G}raph \underline{N}eural \underline{N}etworks model (SDGNN).
In comparison with traditional GNNs, we take related social theories into account and redesign our aggregators and loss functions. 
For signed directed networks, we define four different signed directed relations.
And we propose a layer-by-layer signed relation GNN to aggregate and propagate the information of nodes in signed networks.
For training our model, we reconstruct three important parts of signed directed network: signs, directions, and triangles.

The major contributions of this paper are as follows:
\begin{itemize}
    \item After reviewing the two fundamental social psychology theories of signed network analysis, we conduct an empirical analysis of these theories on five real-world datasets.
    \item We introduce a new layer-by-layer Signed Directed Relation GNN model.
      It aggregates and propagates information between nodes under different signed directed relation definitions.
      Guided by two sociological theories, our loss functions consist of reconstructing signs, directions, and triangles.
    \item We conduct link sign prediction experiments on five real-world signed social network datasets to demonstrate the effectiveness of our proposed model.
\end{itemize}

\section{Related Work}\label{sec:related-work}
\subsection{Signed Network Embedding}
Signed social networks are such social networks in signed social relations having both positive and negative signs\citepp{easley2010networks}.
To mine signed networks, many algorithms have been developed for lots of tasks, such as community detection\citepp{traag2009community}, node classification\citepp{tang2016node}, link prediction\citepp{leskovec2010signed}, spectral graph analysis\citepp{li2016spectral}, and group partition\citepp{huang2018computing}.
Recently, with the development of network representation learning\citepp{perozzi2014deepwalk,grover2016node2vec,tang2015line}, researchers begin to learn low-dimensional representations for signed networks.
For signed network embedding methods, SNE\citepp{yuan2017sne} adopts the log-bilinear model and incorporates two signed-type vectors to capture the positive or negative relationship of each edge along the path.
SiNE\citepp{wang2017signed} designs an objective function guided by social theories to learn signed network embeddings.
SiNE proposes that social theories can provide a fundamental understanding of signed social networks.
For directed signed network, SIDE\citepp{kim2018side} provides a linearly scalable method that leverages balance theory along with random walks. 
SIGNet\citepp{islam2018signet} combines balance theory with specialized random and new sampling techniques in directed signed networks. 
BESIDE \citepp{chen2018bridge} mathematically models ``bridge'' edges based on balance and status theory and achieves state-of-the-art performances.

These methods are devoted to defining an objective function that incorporates sociological theory and then using machine learning techniques (\eg sampling and random walks) to optimize look-up embeddings.

\subsection{Graph Neural Networks}
Today's GNNs can be summarized as Message Passing Neural Networks (MPNNs), including message functions and vertex update functions\citepp{gilmer2017neural}.
Because of the non-Euclidean data structure of the graph, traditional RNNs and CNNs are not easy to be used in the graph domains.
By using GNNs, researchers have successfully applied convolution\citepp{kipf2016semi}, attention\citepp{velickovic2017graph}, LSTM\citepp{hamilton2017inductive}, and other mechanisms into the graph data.
Specifically, Graph Auto-Encoders\citepp{kipf2016variational,pan2018adversarially} are a family of models aimed at mapping (encoding) each node to low-dimensional vectors which reconstructing (decoding) the graph.
Compared to the network embedding methods, GNNs have a partial intersection but use the deep learning methods instead of matrix factorization and random walk and can better describe the network structure and node characteristics\citepp{wu2019comprehensive}. 
A lot of GNN models show a better performance than the shadow look-up embeddings\citepp{kipf2016semi,velickovic2017graph,hamilton2017inductive}.

Most Graph Neural Networks\citepp{kipf2016semi,velickovic2017graph,hamilton2017inductive,xu2019graph,xu2020label} are designed for unsigned social networks whose links are only positive. 
How to apply graph neural networks to signed directed networks faces some challenges (\eg how to model the negative links and the directions).
SGCN\citepp{derr2018signed} designs a new information aggregation and propagation mechanism for the undirected signed networks according to balance theory.
SGCN applies a mean-pooling strategy that is close to GraphSAGE\citepp{hamilton2017inductive} to learning node embeddings.
SiGAT \citepp{huang2019signed} introduces GAT\citepp{huang2019signed} to directed signed networks and designs a motif-based graph neural network model based on social theories.
However, SiGAT uses 38 motifs, which is expensive for large graphs, and its objective functions can only model the sign, ignoring other vital features (\eg directions and triangles).

\section{Problem Definition}\label{sec:problem-definition}
We define a signed directed network as $\G=(\V, \E, s)$, where $\V$ is the set of nodes in a graph $\G$, and $\E$ is the edge list with signs $s$ and directions.
$\E$ consist of $\E^+$ and $\E^-$ while $\E^+ \bigcap \E^- = \emptyset$; $\E^+$ and $\E^-$ denote the sets of positive and negative links, respectively.
It can be denoted as the adjacency matrix of the signed network $A$, where $A_{ij} = 1$ means there exists a positive link from $u_i$ to $u_j$, $A_{ij}=-1$ denotes a negative link from $u_i$ to $u_j$, and $A_{ij}=0$ means there is no link from $u_i$ to $u_j$.  
Given a signed graph $\G=(\V, \E, s)$, our purpose is to map the nodes $u \in \V$ to low-dimensional vectors $z_u\in \mathbb{R}^d$ as:
\begin{equation}
 f (A) \rightarrow Z,
\end{equation}
where $Z \in \mathbb{R}^{|\mathcal{V}|\times d}$ is $d$-dimensional representations for the $|\mathcal{V}|$ nodes of the signed network, $f$ is a learned transformation function.

\section{Sociological Theory}\label{sec:theory}

Two sociological theories (\ie Balance Theory and Status Theory) play an essential role in analyzing and modeling signed directed networks \citepp{leskovec2010signed}.
In this section, we will briefly introduce these two theories and compare them in five real-world datasets.

\begin{figure}[!ht]
  \centering
\begin{subfigure}[t]{0.23\textwidth}
   \includegraphics[width=\textwidth]{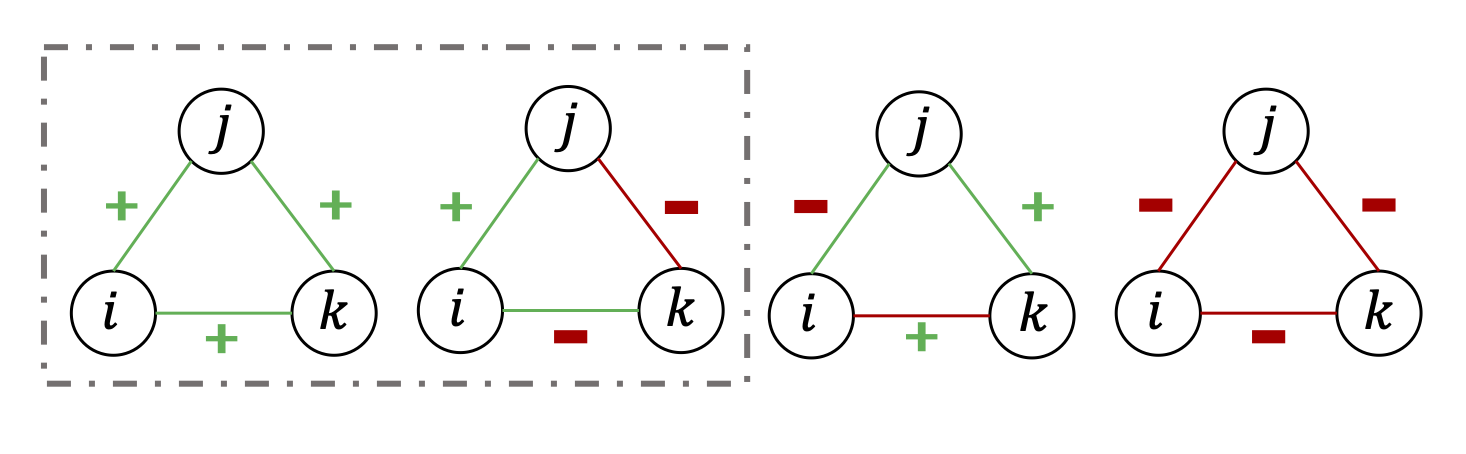}
  \caption{Balance theory} 
  \label{fig:balance_theory}
\end{subfigure}
\begin{subfigure}[t]{0.23\textwidth}
  \includegraphics[width=\textwidth]{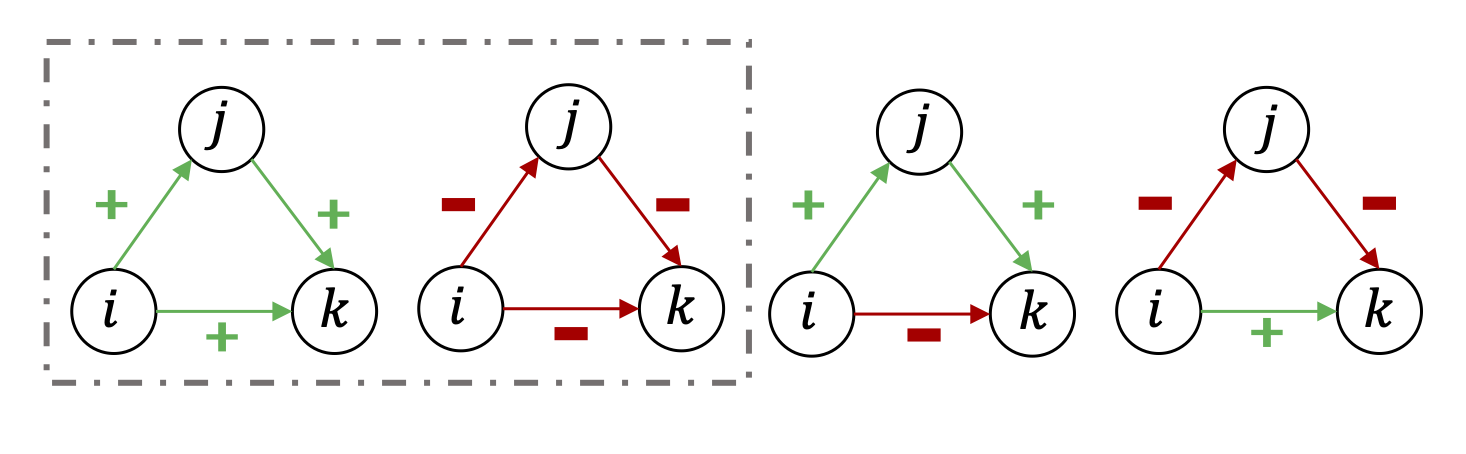}
  \caption{Status theory} 
  \label{fig:status_theory}
\end{subfigure}
\caption{Illustrations of two sociological theories.}
\end{figure}

\subsection{Balance Theory}
Balance theory originated in the 1950s \citepp{heider1946attitudes} is initially intended as a model for undirected signed networks.
Triads with an even number of negative edges as balanced.
In \figref{fig:balance_theory}, the triangles with three positive signs and those with one positive sign (\ie the first two triads in \figref{fig:balance_theory}) are balanced.
Balance theory posits that balanced triads are more plausible — and hence should be more prevalent in real-world networks — than unbalanced triads.
It exemplifies the principle that \textit{the friend of my friend is my friend} and \textit{the enemy of my enemy is my friend}.
Balance theory is widely used in the field of signed networks \citepp{leskovec2010predicting}.

\subsection{Status Theory}
Status theory is another critical sociological theory in signed network analysis, which provides a different organizing principle for directed networks of signed links.
It supposes that a positive directed link ``+'' indicates that the creator of the link views the recipient as having higher status; and a negative directed link ``-'' indicates that the recipient is viewed as having lower status \citepp{leskovec2010signed}.
The status may denote the relative prestige, ranking, or reputation.
For example, a positive link from A to B means not only \textit{B is A's friend} but also \textit{B has a higher status than A}.
For the triangles in \figref{fig:status_theory}, the first two triads satisfy the status order, but the last two do not satisfy it.
For the first triads, when $\mathrm{Status(j)} > \mathrm{Status(i)}$ and $\mathrm{Status(k)} > \mathrm{Status(j)}$, we have $\mathrm{Status(k)} > \mathrm{Status(i)}$. 

\subsection{Comparison of Balance Theory and Status Theory}

Balance theory focuses on undirected signed networks, although it has been applied to directed networks by simply disregarding the directions \citepp{wasserman1994social}.
While, status theory normally reflects relations between two users, which is based on directions.
In some cases, these two theories can be consistent with predictions.
\begin{table}[!ht]
  \centering
  \scalebox{0.9}{
    \begin{tabular}{cccccc}
      \hline
      Dataset  & Both & Only Balance & Only Status & Neither  \\
      \hline
      Bitcoin-Alpha   & 0.673 & 0.208 & 0.094 & 0.025 \\
      Bitcoin-OTC     & 0.686 & 0.208 & 0.083 & 0.023 \\
      Wikirfa         & 0.686 & 0.059 & 0.189 & 0.066 \\
      Slashdot        & 0.751 & 0.167 & 0.066 & 0.016  \\
      Epinions        & 0.769 & 0.156 & 0.066 & 0.009  \\
      \hline
    \end{tabular}
  }
  \caption{The ratio of triads satisfying balance and/or status theory.}
  \label{tb:two-theory-dataset}
  \vspace{-10px}
\end{table}
This phenomenon has received a lot of attention from researchers \citepp{leskovec2010signed,chen2018bridge}.

We follow the related works and examine the percentage of triads satisfying balance and/or status theory on five real-world datasets used in this paper.

From \tableref{tb:two-theory-dataset}, we can find that only a tiny fraction of triangles satisfies neither of two theories.
About 70\% of triads can be consistent with both theories.
We think these triangles that satisfy neither theory maybe the noise for the signed network embedding representation.
In addition, balance theory models the relationship among three vertices, and status theory capture the relationship between two vertices with the transitivity property \citepp{chen2018bridge}.
The complementation of both theories can be the key point for signed directed network representation learning.

\section{Proposed Method}\label{sec:methods}
Based on the previous discussion on sociological theory in signed directed networks, we will introduce how to design our new SDGNN model in this section.

\subsection{Signed Directed GNNs}

For unsigned networks, neighborhood nodes have the same semantic relations.
GraphSAGE\citepp{hamilton2017inductive} uses a mean aggregator to aggregate information from neighborhoods.
As the weights in GraphSAGE only determined by network structure, GAT\citepp{velickovic2017graph} learns the weights by structure masked self-attention mechanism.
We adapt the above GNNs aggregators to signed directed networks.

\begin{figure*}[!htp]
  \centering
  \includegraphics[width=0.7\textwidth]{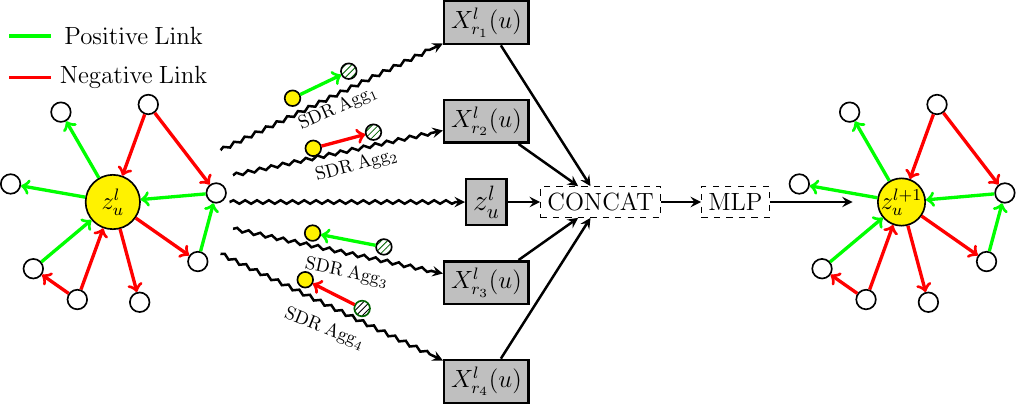}
  \caption{An Illustration of SDGNN model. We use Signed Directed Relation (SDR) aggregators to aggregate and propagate the information of nodes. }
  \label{fig:sgae-model}
  \vspace{-15px}
\end{figure*}

\figref{fig:sgae-model} depicts the overall architecture of the proposed SDGNN model.
For a signed directed graph, the directions and signs between nodes reflect different relations and semantics.
Different neighborhoods with different relations should be distinguished.
We first define 4 different signed directed relation  (\ie $ u\rightarrow^+ v,u\rightarrow^- v, u\leftarrow^+ v, u\leftarrow^- v$).
Based on a signed directed relation (SDR) ${r_i}$, we can get the neighborhoods $\mathcal{N}_{r_i}$.
After that, we use different SDR GNN aggregators to aggregate the information from different neighborhoods and use an MLP to encode these messages into node embeddings.
GNN aggregators can be mean aggregators or attention aggregators.
For a mean aggregator, we get the information $X^l_{r_i}(u)$ and concatenate  $z_u^l$ with $X^l(u)$  by:
\begin{equation}
  \begin{aligned}
  X_{r_i}^{l}(u) & = \sigma (\mathbf{W}^l_{r_i} \cdot \mathrm{MEAN} (\{z_u^l\} \cup \{z_v^l, \forall v \in \mathcal{N}_{r_i}(u) \})) \\
  z^{l+1}_u & = \mathrm{MLP} (\mathrm{CONCAT} (X^l(u), X_{r_1}^l(u),..., X_{r_i}^l(u))),
  \end{aligned}
\end{equation}
where $z_u$ is the embedding of node $u$, $\mathcal{N}_{r_i}(u)$ is the neighborhoods of $u$ under the definition $r_i$, $\mathbf{W}$ is the parameter, and $\sigma$ is the activation function.

For attention aggregators, we first compute  $\alpha^{r_i}_{uv}$ for the node $u$ and node $v$ by the attention mechanism $\vec{\mathbf a}_{r_i}$ and LeakyReLU nonlinearity activation function (with negative input slope $\alpha$ = 0.2) as:
  \begin{equation}
    \begin{aligned}
      \label{eq:gat1}
      \alpha_{uv}^{r_i}  = \frac{\exp\left(\text{LeakyReLU}\left(\vec{\bf a}_{r_i}^\top[{\bf W}^l_{r_i}{z^l_u}\|{\bf W}^l_{r_i}{z^l_v}]\right)\right)}{\sum_{k\in\mathcal{N}_{r_i}(u)} \exp\left(\text{LeakyReLU}\left(\vec{\bf a}_{r_i}^\top[{\bf W}^l_{r_i}{z^l_u}\|{\bf W}^l_{r_i}{z^l_k}]\right)\right)}, \\
    \end{aligned}
  \end{equation}
where $\|$ is the concatenation operation, $\mathcal{N}_{r_i}(u)$ is the neighborhoods of node $u$ under the definition of $r_i$, $\mathbf W_{r_i}$ is the weight matrix parameter.
Then we get the information $X^l_{r_i}(u)$ and concatenate  $X^l_{r_i}(u)$ with $z_u$ to get new embeddings:

  \begin{equation}
    \begin{aligned}
      \label{eq:gat2}
      X^l_{r_i}(u) &  = \sum_{v\in\mathcal{N}_{r_i}(u)} \alpha_{uv}^{r_i} {\mathbf W}_{r_i}z^l_v, \\
       z^{l+1}_u & = \mathrm{MLP} (\mathrm{CONCAT} (X^l(u), X^l_{r_1}(u),..., X^l_{r_i}(u))).
    \end{aligned}
  \end{equation}

In \figref{fig:sgae-model}, we use a layer-by-layer Signed Directed GNN to cover the direction and the topology of the triangle when we use multi-layers instead of one layer.
  
\subsection{Loss Function}
For training our SDGNN model, we design three loss functions to reconstruct three critical features of signed networks: sign, direction, and triangle.

For sign, we use the following cross entropy loss function to model the sign between two nodes:  
{\small
  \begin{equation}\label{eq:loss-function}
    \begin{aligned}
      s_{u,v} = & \sigma(z^\top_uz_v) \\
      \mathcal{L}_{sign}(u, v) = & - y_{uv} \log\left( s_{u, v}\right) 
      - (1-y_{uv})\log\left(1 - s_{u, v}\right), \\
      \mathcal{L}_{sign} = & \sum\limits_{e_{u,v} \in \E} \mathcal{L}_{sign} (u, v),
    \end{aligned}
  \end{equation}
}%
where $\sigma$ is the sigmoid function, $y_{uv}$ is the sign ground truth, and $\E$ is the edge list with signs.

As we discussed before, status theory relies on the directions in signed networks.
We denote the status ranking score of node $u$, $v$ as $s(z_u)$, $s(z_v)$.
We use the following square loss function to measure the difference between the predicted status relationship value $s(z_u) - s(z_v)$ from the edge $e_{uv}$ and the ``ground truth'' value $q_{uv}$:
\begin{equation}
  \begin{aligned}
  \mathcal{L}_{direction}(u \rightarrow v)  & =  (q_{uv} - (s(z_u) - s(z_v)))^2 \\
  q_{uv} & = 
  \begin{cases}
    \mathrm{max}(s(z_u) - s(z_v), \ \gamma) &  u \rightarrow v: - \\
    \mathrm{min}(s(z_u) - s(z_v), -\gamma) & u \rightarrow v: + \\
  \end{cases} \\
  \mathcal{L}_{direction} & = \sum\limits_{e_{u,v} \in \E} \mathcal{L}_{direction} (u \rightarrow v),
\end{aligned}
\end{equation}
where $s(z) = \mathrm{sigmoid}(W \cdot z + b)$ is a score function for mapping embedding $z$ to a score, $\gamma$ is a threshold of status relationship value ($\gamma=0.5$ in this paper), and $\E$ is the edge list with signs. 

For triangles, we hope our model can learn from the true triangles distribution. 
For a triangle (\eg $\bigtriangleup_{i,j,k}, i \rightarrow^+ j, i \rightarrow^+ k, k \rightarrow^+ j$),  we maximize the likelihood by:

\begin{equation}
  \begin{aligned}
  J_{\bigtriangleup_{i,j,k}} & = P(+|e_{ij}) * P(+|e_{ik}) * P(+|e_{kj}). \\
\end{aligned}
\end{equation}

Further, triangles objective function $J_{tri}$ as follows:
  \begin{equation}
    \begin{aligned}
      \label{eq:tri0}
  J_{tri} & = \prod_{\bigtriangleup \in T} J_{\bigtriangleup},  \\
  \mathcal{L}_{triangle} & = - \mathrm{log} J_{tri} = \sum\limits_{\triangle \in T} -\mathrm{log} J_{\bigtriangleup} = \sum\limits_{\bigtriangleup \in T} \mathcal{L}_{\bigtriangleup},
  \end{aligned}
\end{equation}
where $T$ is the set of triangles based on balanced and status theories.
We can construct a triangle by three vertices ($i,j,k$) by:
{\small
\begin{equation}
  \label{eq:tri1}
  \begin{aligned}
    \mathcal{L}_{\bigtriangleup_{i,j,k}} & = \mathcal{L}_{ij} + \mathcal{L}_{ik} + \mathcal{L}_{kj} ,\\
    \mathcal{L}_{ij} & = -y_{i,j} \mathrm{log}P(+|e_{ij}) - (1-y_{i,j}) \mathrm{log} (1-P(+|e_{ij})) \\
    & = -y_{i,j} \log\left(\sigma(z^\top_iz_j) \right) - (1-y_{i,j})\log( 1 - \sigma(z_i^\top z_j)),
   \end{aligned}
 \end{equation}
}
 where $y_{i,j}$ is the sign ground truth for edge $e_{ij}$.
 With \equref{eq:tri0} and \equref{eq:tri1}, we can reconstruct triangles by edge binary cross entropy loss function.
 The weight of the edge loss function is the number of edges in triangles.

 Based on sign, direction and triangle loss function, the overall objective function is written as:
 \begin{equation}
   \label{eq:loss}
   \mathcal{L}_{loss} = \mathcal{L}_{sign} + \lambda_{1} \mathcal{L}_{direction} + \lambda_{2} \mathcal{L}_{triangle},
 \end{equation}
 where $\lambda_1$ and $\lambda_2$ are the weight of different loss functions.
 \equref{eq:loss} shows that our loss functions are designed to reconstruct the various properties of signed networks.

\section{Experiments}\label{sec:experiments}
In this section, we conduct link sign prediction to check whether our model improves the performance of signed network embeddings.
Link sign prediction is the task of predicting the unobserved sign of existing edges in the test dataset given training dataset\citepp{leskovec2010predicting,chen2018bridge}.
We follow their experimental settings and compare our method against some state-of-the-art embedding methods.

\subsection{Experimental Settings}

\subsubsection{Datasets}

We do experiments on five real-world signed social network datasets (\ie Bitcoin-Alpha\footnote{http://snap.stanford.edu/data/soc-sign-bitcoin-alpha.html}, Bitcoin-otc\footnote{http://snap.stanford.edu/data/soc-sign-bitcoin-otc.html}, Wikirfa\footnote{http://snap.stanford.edu/data/wiki-RfA.html},
Slashdot\footnote{http://snap.stanford.edu/data/soc-sign-Slashdot090221.html} and Epinions\footnote{http://snap.stanford.edu/data/soc-sign-epinions.html}). 
Bitcoin-Alpha and Bitcoin-OTC\citepp{kumar2016edge} are the who-trust-whom networks of people who trade using Bitcoin on platforms.
In these datasets, members rate other members on a scale of -10 (total distrust) to +10 (total trust) in steps of 1.
We treat the scores greater than 0 as positive and others as negative. 
Wikirfa\citepp{kumar2016edge} is a signed network in which nodes represent Wikipedia members and edges represent votes.
It records the voting results for ``request for adminship (RfA)'', where the community member can cast a supporting, neutral, or opposing vote for a Wikipedia adminship election.
We remove the neural votes and construct a signed network \citepp{chen2018bridge}.
Slashdot\citepp{leskovec2010signed} is from a technology-related news website with user communities.
The website introduced Slashdot Zoo features that allow users to tag each other as friends or foes.
The dataset is a common signed social network with friends and enemies labels.
Epinions\citepp{leskovec2010signed} is a who-trust-whom online social network of a consumer review site Epinions.com.
Members of the site can indicate their trust or distrust of the reviews of others.
The network reflects people's opinions on others.

\begin{table}[!ht] \centering
  \scalebox{0.9}{
  \begin{tabular}{cccccc}
    \hline
    Dataset  & \# nodes & \# pos links & \# neg links & \% pos   \\
    \hline
    Bitcoin-Alpha & 3,783 & 22,650 & 1,536 & 93.65 \\
    Bitcoin-OTC & 5,881 & 32,029  & 3,563 & 89.99\\
    Wikirfa & 11,259 & 138,813  & 39,283 & 77.94 \\
    Slashdot & 82,140   & 425,072      & 124,130      & 77.40  \\
    Epinions & 131,828   & 717, 667  & 123,705      & 85.30    \\
    \hline
  \end{tabular}
  }
  \caption{Statistics of five datasets.}
  \label{tb:dataset}
\end{table}

The statistics of five datasets are summarized in Table~\ref{tb:dataset}.
For these five datasets, positive and negative links are imbalanced (\ie nearly 80\% links are positive edges). 

\subsubsection{Baselines}

To validate the effectiveness of SDGNN, we compare it with a number of baselines including unsigned network embedding methods, signed embedding methods and sigend graph neural networks.
\begin{itemize}
    \item Random: It generates $d$ dimensional random values, $z = (x_1, x_2, ...,x_{d}), x_i \in [0.0, 1.0)$. It can be used to show the ability of the downstream logistic regression.

    \item Unsigned Network Embedding: We use several classical unsigned network embedding methods to validate the effectiveness of the structure: DeepWalk\citepp{perozzi2014deepwalk}, Node2vec\citepp{grover2016node2vec}, and LINE\citepp{tang2015line}. 
    Since these methods cannot distinguish positive and negative edges, we remove the negative links in the training stage.

    \item Signed Network Embedding: In order to show the effectiveness of modeling signed directed edges, we use some signed network embedding methods (\eg, SiNE\citepp{wang2017signed}, SIGNet\citepp{islam2018signet}, BESIDE\citepp{chen2018bridge}).
      Specifically, SiNE is designed for undirected signed networks. 
      SIGNet and BESIDE can model signed directed networks.
      
    \item FeExtra\citepp{leskovec2010predicting}: This method extracts two parts, a total of 23 features from signed directed social network. For each pair ($v_i$,$v_j$), the first one is the degree based, such as the number of incoming positive and negative links of $v_i$, the number of outgoing positive and negative links of $v_j$ and so on. The other one is the structure information of the triad that contains $v_i$ and $v_j$.

    \item Graph Neural Network: For the baselines of GNNs, we choose SGCN\citepp{derr2018signed} and SiGAT\citepp{huang2019signed}. SGCN makes a dedicated and principled effort that utilizes balance theory to correctly aggregate and propagate the information across layers of a undirected signed GCN model. SiGAT uses 38 motifs and GAT aggregators to model signed directed networks.

\end{itemize}

\begin{table*}[!ht]
  \centering 
  \scalebox{0.9}{
    \setlength{\tabcolsep}{1.0mm}{
      \begin{tabular}{c|c|c|ccc|ccc|c|ccc}
    \toprule
    \multicolumn{2}{c|}{} & \multicolumn{1}{c|}{Random}  
    & \multicolumn{3}{c|}{\begin{tabular}[c]{@{}c@{}}Unsigned \\Network Embedding\end{tabular}} 
    & \multicolumn{3}{c|}{\begin{tabular}[c]{@{}c@{}}Signed \\Network Embedding\end{tabular}} 
    & \multicolumn{1}{c|}{\begin{tabular}[c]{@{}c@{}}Feature\\ Engineering\end{tabular}}
    & \multicolumn{3}{c}{\begin{tabular}[c]{@{}c@{}}Graph \\Neural Network\end{tabular} } \\
    \midrule
    Dataset                     & Metric   & Random & Deepwalk & Node2vec & LINE & SiNE  & SIGNet  & BESIDE & FeExtra  & SGCN & SiGAT & SDGNN  \\
    \midrule
\multirow{4}{*}{Bitcoin Alpha} & Micro-F1 & 0.9367 &  0.9367 &  0.9355 &  0.9352 &  0.9458 &  0.9422   &  \underline{0.9489} &  0.9486 &  0.9256 &  0.9456 &  \textbf{0.9491} \\
                               & Binary-F1 & 0.9673 &  0.9673 &  0.9663 &  0.9664 &  0.9716 & 0.9696  &  \textbf{0.9732} &  \underline{0.9730} &  0.9607 &  0.9714 &  0.9729 \\
                               & Macro-F1 & 0.4837 &  0.4848 &  0.6004 &  0.5220 &  0.6869 & 0.6965  &  \underline{0.7300} &  0.7167 &  0.6367 &  0.7026 &  \textbf{0.7390} \\
                          & AUC & 0.6146 &  0.6409 &  0.7576 &  0.7114 &  0.8728 & 0.8908  &  \underline{0.8981} &  0.8882 &  0.8469 &  0.8872 &  \textbf{0.8988} \\
    \midrule
\multirow{4}{*}{Bitcoin-OTC}   & Micro-F1 & 0.9000 &  0.8937 &  0.9089 &  0.8911 &  0.9095 & 0.9229  &  0.9320 &  \textbf{0.9361} &  0.9078 &  0.9268 &  \underline{0.9357} \\
                               & Binary-F1 & 0.9473 &  0.9434 &  0.9507 &  0.9413 &  0.9510 & 0.9581  &  0.9628 &  \textbf{0.9653} &  0.9491 &  0.9602 &  \underline{0.9647} \\
                      & Macro-F1 & 0.4737 &  0.5281 &  0.6793 &  0.5968 &  0.6805 & 0.7386  &  \underline{0.7843} &  0.7826 &  0.7306 &  0.7533 &  \textbf                                                                                                                                   {0.8017} \\
                          & AUC & 0.6145 &  0.6596 &  0.7643 &  0.7248 &  0.8571 & 0.8935  &  \textbf{0.9152} &  0.9121 &  0.8755 &  0.9055 &   \underline{0.9124} \\
    \midrule
\multirow{4}{*}{Wikirfa} & Micro-F1 & 0.7797 &  0.7837 &  0.7814 &  0.7977 &  0.8338 & 0.8384  &  \underline{0.8589} &  0.8346 &  0.8489 &  0.8457 &  \textbf{0.8627} \\
                               & Binary-F1 & 0.8762 &  0.8779 &  0.8719 &  0.8827 &  0.8972 & 0.9001  &  \underline{0.9117} &  0.8987 &  0.9069 &  0.9042 &  \textbf{0.9142} \\
                               & Macro-F1 & 0.4381 &  0.4666 &  0.5626 &  0.5738 &  0.7319 & 0.7384  &  \underline{0.7803} &  0.7235 &  0.7527 &  0.7535 &  \textbf{0.7849} \\
                          & AUC & 0.5423 &  0.5876 &  0.6930 &  0.6772 &  0.8602 & 0.8682  &  \textbf{0.8981} &  0.8604 &  0.8563 &  0.8829 &  \underline{0.8898} \\
    \midrule
\multirow{4}{*}{Slashdot} & Micro-F1 & 0.7742 &  0.7738 &  0.7526 &  0.7489 &  0.8265 & 0.8389  &  \underline{0.8590} &  0.8472 &  0.8296 &  0.8494 & \textbf{0.8616} \\
                               & Binary-F1 & 0.8728 &  0.8724 &  0.8528 &  0.8525 &  0.8918 & 0.8983   &  \underline{0.9105} &  0.9070 &  0.8926 &  0.9055 &  \textbf{0.9128}\\
                               & Macro-F1 & 0.4364 &  0.4384 &  0.5390 &  0.5052 &  0.7273 & 0.7554  &  \textbf{0.7892} &  0.7399 &  0.7403 &  0.7671 &  \textbf{0.7892}\\
                          & AUC & 0.5370 &  0.5408 &  0.6709 &  0.6145 &  0.8409 & 0.8752  &  \textbf{0.9017} &  0.8880 &  0.8534 &  0.8874 & \underline{0.8977} \\
    \midrule
\multirow{4}{*}{Epinions} & Micro-F1 & 0.8525 &  0.8214 &  0.8563 &  0.8535 &  0.9173 & 0.9113  &  \underline{0.9336} &  0.9226 &  0.9112 &  0.9293 & \textbf{0.9355}\\ 
                               & Binary-F1 & 0.9204 &  0.9005 &  0.9170 &  0.9175 &  0.9525 & 0.9489  &  \underline{0.9615} &  0.9561 &  0.9486 &  0.9593 & \textbf{0.9628}\\  
                               & Macro-F1 & 0.4602 &  0.5131 &  0.6862 &  0.6305 &  0.8160 & 0.8060  &  \underline{0.8601} &  0.8130 &  0.8105 &  0.8454 &  \textbf{0.8610}\\
                          & AUC & 0.5589 &  0.6702 &  0.8081 &  0.6835 &  0.8872 & 0.9095  &  0.9351 &  \textbf{0.9444} &  0.8745 &  0.9333 &  \underline{0.9411}\\
    \bottomrule

\end{tabular}
    }
}
\caption{The results of link sign prediction on five datasets.}
\label{tb:experiment-result}
\vspace{-14px}
\end{table*}
For a fair comparison, the embedding dimension $d$ is set to 20 for all methods except FeExtra.
It is a common setting used in SiNE\citepp{wang2017signed}, BESIDE\citepp{chen2018bridge} and SiGAT\citepp{huang2019signed}.
We use the authors' released code for DeepWalk\footnote{https://github.com/phanein/deepwalk}, Node2vec\footnote{https://github.com/aditya-grover/node2vec}, LINE\footnote{https://github.com/tangjianpku/LINE}, SiNE\footnote{http://www.public.asu.edu/\textasciitilde swang187/codes/SiNE.zip},   SIGNet\footnote{https://github.com/raihan2108/signet}, BESIDE\footnote{https://github.com/yqc01/BESIDE}, and SiGAT\footnote{https://github.com/huangjunjie95/SiGAT}. For SGCN, we use the code from github\footnote{https://github.com/benedekrozemberczki/SGCN}. 
We follow the authors' suggested hyperparameter settings.
Like previous works\citepp{kim2018side,wang2017signed,derr2018signed}, we first use these methods to get node representations. 
For edge $e_{ij}$, we concatenate these two learned representation $z_i$ and $z_j$ to compose an edge representation $z_{ij}$. 
After that, we train a logistic regression classifier on the training set and use it to predict the edge sign in the test set. 
We randomly select 80\% edges as training set and the remaining 20\% as the test set.
We run with different train-test splits for 5 times to get the average scores.
Each training set is used to train both embedding vectors and logistic
regression classifiers.
Our models are implemented by PyTorch with the Adam optimizer ($\mathrm{LearningRate}=0.001$, $\mathrm{WeightDecay} = 0.001$). 
We use the 2-layer-GAT aggregators to build our model and set $\lambda_1=1$ and $\lambda_2=1$.
All experiments run on a computer with Intel Xeon E5-2640 CPU and 128GB RAM, which installs Linux CentOS 7.1.

  \begin{table*}[ht]
    \centering
    \scalebox{0.9}{
      \begin{tabular}{cccccc}
        \toprule
        Metric  & 2-Layer-GAT-AGG & 1-Layer-GAT-AGG & 2-Layer-MEAN-AGG & 1-Layer-MEAN-AGG  \\
        \midrule
        Micro-F1 & 0.9446 & 0.9442 & 0.9415 & 0.9399\\
        Binary-F1 & 0.9706 & 0.9703 & 0.9689 & 0.9679\\
        Macro-F1 & 0.7510 & 0.7516 & 0.7417 & 0.7411\\
        AUC  & 0.9154 & 0.9095 & 0.9041 & 0.9000\\
        \bottomrule
      \end{tabular}
    }
     \caption{Ablation study on different aggregators.}
    \label{tb:GNN-aggregator}
    \vspace{-8px}
  \end{table*}

  \begin{table*}[!ht]
    \centering
    \scalebox{0.9}{
      \begin{tabular}{c|ccccc}
        \toprule
        Metric  & $\mathcal{L}_{sign}$  & $\mathcal{L}_{sign}+\mathcal{L}_{direction}$ & $\mathcal{L}_{sign}+\mathcal{L}_{triangle} $ &
                                                                                                                                        $\mathcal{L}_{sign}+\mathcal{L}_{direction}+\mathcal{L}_{triangle}$ \\
        \midrule
        Micro-F1 & 0.9386 & 0.9438 & 0.9415 & 0.9475 \\
        Binary-F1 & 0.9677 & 0.9702 & 0.9690 & 0.9721 \\
        Macro-F1 & 0.6738 & 0.7414 & 0.7210 & 0.7585 \\
        AUC &  0.8883 & 0.9082 & 0.9030 & 0.9109 \\
        \bottomrule
      \end{tabular}
    }
    \caption{Ablation study on loss functions.}
    \label{tb:ablation-decoder}
    \vspace{-14px}
  \end{table*}

\subsection{Experiment Results}

We report the average Micro-F1, Binary-F1, Macro-F1, and AUC in Table~\ref{tb:experiment-result} \citepp{scikit-learn}. 
We have bolded the highest value of each row and underlined the second value. From Table~\ref{tb:experiment-result}, we can find that:
\begin{itemize}
    \item For signed networks, link sign prediction is a positive and negative imbalance classificiation problem.
    Even given random embedding, logistic regression can be used as the downstream machine learning methods.

  \item After using unsigned network embedding methods, even the only positive links are used, the metrics have been increased.
    It means that the structural information matters.
    Node2vec has made the best results in the unsigned network embeddings methods.
    
  \item Signed network embedding methods (\ie SiNE, SIGNet, and BESIDE) are designed for signed network.
    The results are significantly higher than other unsigned network embedding methods.
    These algorithms model related sociological theories and had achieved good results for signed network analysis.
    SiNE is designed for undirected signed networks based on balance theory.
    Its results are not as good as SIGNet and BESIDE.
    Although SIGNet can be applied to signed directed networks, it does not model status theory.
    This makes it is less effective than BESIDE.
    BESIDE shows some good results in our experiments, which demonstrates that modeling two theories is the key for the problem.
  \item FeExtra method is a very classic method in the early days and performs well because of its successful use of relevant sociological theories.
    However, it should be pointed out that this method relies on feature engineering to extract features manually and only model edge features without node embeddings, so its generalization ability is weak.
  \item SGCN shows a performance close to SiNE, but it cannot effectively model signed directed networks because the algorithm did not consider the direction.
    Besides, it uses the $\mathrm{MEAN}$ aggregators which can not model the effects of different neighborhoods. SiGAT is designed for directed networks and uses the attention aggregators. Its experimental results are better than SGCN. But its decoder design limits its expressiveness.
  \item  SDGNN outperforms all baseline methods in most terms of metrics.
  It demonstrates the ability of learning node embeddings using our SDGNN model.
  It is worth mentioning that the experimental results depend on the downstream classifier, and there may exist some inconsistent between F1 and AUC.
  \end{itemize}

  \subsection{Parameter Analysis and Ablation Study}

  In this section, we investigate the effects of hyperparameters and do some ablation studies to analyze our model architecture design.
We choose the Bitcoin-Alpha as our dataset and select 80\% training edges and 20\% test edges as the previous subsection does. We also use logistic regression function as our downstream machine learning classifier.

\subsubsection{Parameter Analysis}

In this subsection, we analyze the hyperparameters about the number of epoch and the embedding dimension $d$.
\begin{figure}[!ht]
  \centering
  \begin{subfigure}[t]{0.52\linewidth}
    \includegraphics[width=1.05\linewidth]{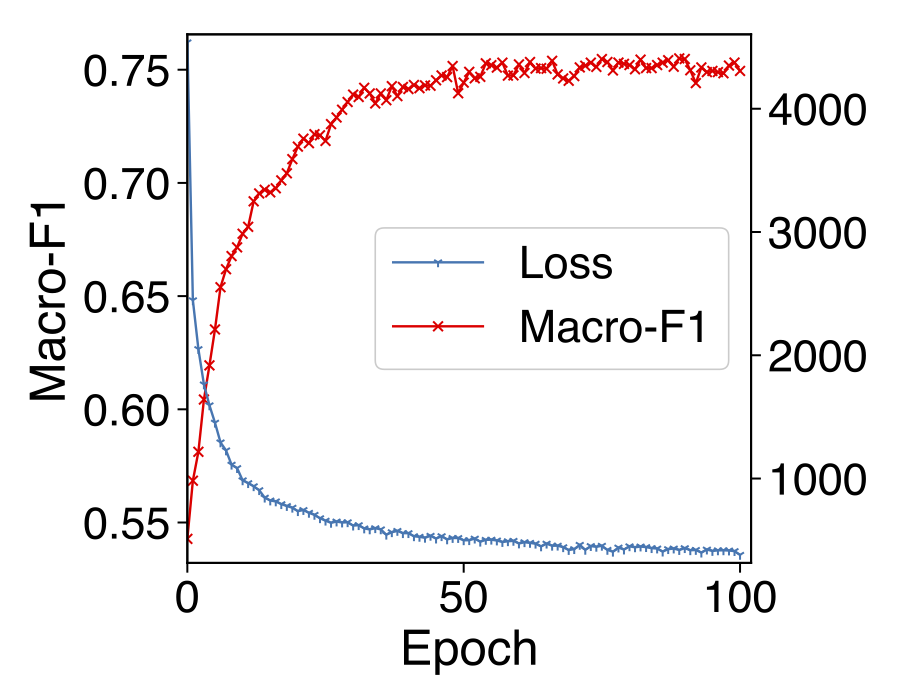}
    \caption{Epoch}
    \label{fig:epoch}
  \end{subfigure}
  \begin{subfigure}[t]{0.47\linewidth}
    \includegraphics[width=\linewidth]{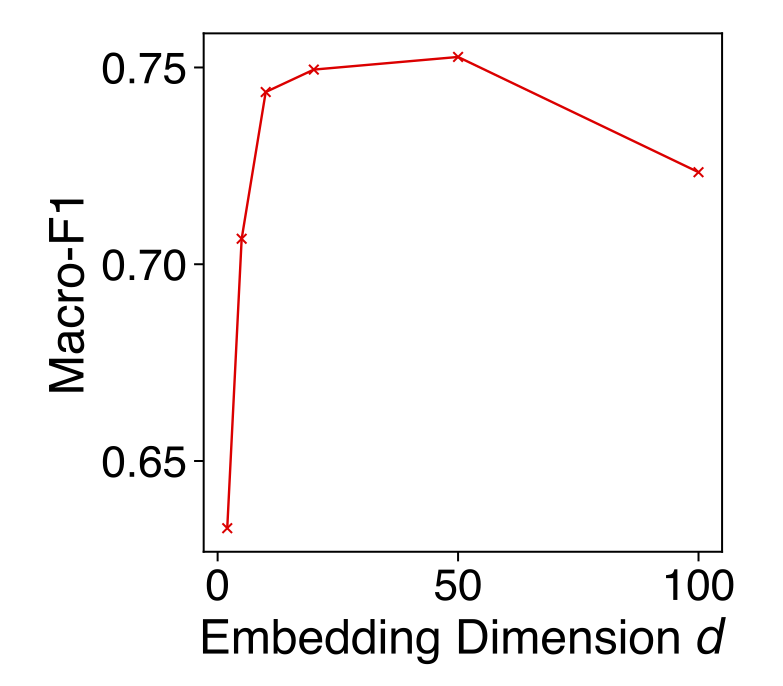}

    \caption{$d$}
    \label{fig:d}
  \end{subfigure}
  \caption{Parameter analysis for sign prediction.}
  \label{fig:parameter-analysis}
  \vspace{-8px}
 \end{figure}

 When analyzing epoch, we set $d=20$, and record the Macro-F1 performance of different epoch generated representations and the loss value during training.
 \figref{fig:epoch} shows that SDGNN converges quickly and keeps relatively stable performances.
 When discussing the dimension $d$, we set the corresponding Epoch to 100 to discuss the robustness of different dimensions.
 \figref{fig:d} shows that the performance increases first and decreases in 100 (We decrease the learning rate for the large $d$).
 Even with a small embedding dimension like $d=5$, our model has already achieved pretty good performance which is close to BESIDE\citepp{chen2018bridge}.
 We find that too many parameters will cause the difficulties of training embeddings, which may be the reason for the decrease in large dimensions.

\subsubsection{Ablation Study}

In this subsection, we do some ablation studies to discuss the design of aggregators and loss functions.
As the same as previous subsection, we choose Bitcoin-Alpha as the dataset to analyze different architecture designs.
For the aggregators, as we mentioned, mean aggregator and attention aggregator are common GNNs models. 
We analyze whether the attention mechanism can perform better than mean strategy. 
In addition, we also discusse the number of GNN layers to verify our layer-by-layer design.
\tableref{tb:GNN-aggregator} demonstrates that the attention mechanism performs better than the mean strategy. 
And the number of layers can capture higher-order information, which can increase the performance.

For our loss functions, we can control $\lambda_1$ and $\lambda_2$ in \equref{eq:loss} to discuss the effects of different objectives on the results.
When $\lambda_1=0$ and $\lambda_2=0$, it degenerates to just reconstruct the sign of signed networks, which is close to SiGAT.
In \tableref{tb:ablation-decoder}, we can find that only reconstructing signs using $\mathcal{L}_{sign}$ performs poorly.
When considering direction and triangle, the results are improved. 
It demonstrates that our training objectives should take both directions and triangles into consideration.
Signs, directions, and triangles are vital features for signed directed networks.

\section{Conclusion}\label{sec:conclusion}
In this paper, we investigate the signed directed network representations learning.
We firstly analyze two fundamental social theories (\ie Balance Theory and Status Theory) in the signed directed network.
Guided by sociological theories, we propose SDGNN to encode a signed network into network embeddings.
SDGNN aggregates messages from different signed directed relation definitions.
It can apply multi-layers to capture high order structure information.
To train our SDGNN, we introduce combined loss functions to reconstruct not only the signs but also other important features for signed directed networks (\ie directions and triangles). 
We perform experiments on five real-world signed networks and demonstrate that our proposed SDGNN performs better than other state-of-art baselines.  
We analyze the hyperparameters and do some ablation studies to analyze our design for aggregators and loss functions.
In future work, we will further generalize this method to heterogeneous networks to incorporate more complex semantic information\citepp{schlichtkrull2018modeling}.

\section*{Acknowledgments}
The authors would like to thank the AAAI reviewers for their insightful suggestions to improve the manuscript.
This work is funded by the the National Key R\&D Program of China 2020AAA0105200 and National Natural Science Foundation of China under Grant No. 91746301 and 61802371. This work is supported by Beijing Academy of Artificial Intelligence (BAAI) under Grant No. BAAI2019QN0304. Huawei Shen is also funded by K.C. Wong Education Foundation.

\bibliographystyle{aaai21}
\bibliography{refs}

\begin{thebibliography}{37}
\providecommand{\natexlab}[1]{#1}
\providecommand{\url}[1]{\texttt{#1}}
\providecommand{\urlprefix}{URL }
\expandafter\ifx\csname urlstyle\endcsname\relax
  \providecommand{\doi}[1]{doi:\discretionary{}{}{}#1}\else
  \providecommand{\doi}{doi:\discretionary{}{}{}\begingroup
  \urlstyle{rm}\Url}\fi

\bibitem[{Chen et~al.(2018{\natexlab{a}})Chen, Qian, Liu, and
  Sun}]{chen2018bridge}
Chen, Y.; Qian, T.; Liu, H.; and Sun, K. 2018{\natexlab{a}}.
\newblock " Bridge" Enhanced Signed Directed Network Embedding.
\newblock In \emph{CIKM}, 773--782.

\bibitem[{Chen et~al.(2018{\natexlab{b}})Chen, Qian, Zhong, and
  Li}]{chen2018bassi}
Chen, Y.; Qian, T.; Zhong, M.; and Li, X. 2018{\natexlab{b}}.
\newblock BASSI: Balance and Status Combined Signed Network Embedding.
\newblock In \emph{DASFAA}, 55--63. Springer.

\bibitem[{Cui et~al.(2020)Cui, Zhuang, Liu, and Wang}]{cui2020semi}
Cui, J.; Zhuang, H.; Liu, T.; and Wang, H. 2020.
\newblock Semi-Supervised Gated Spectral Convolution on a Directed Signed
  Network.
\newblock \emph{IEEE Access} 8: 49705--49716.

\bibitem[{Derr, Ma, and Tang(2018)}]{derr2018signed}
Derr, T.; Ma, Y.; and Tang, J. 2018.
\newblock Signed graph convolutional networks.
\newblock In \emph{ICDM}, 929--934. IEEE.

\bibitem[{Duran and Niepert(2017)}]{duran2017learning}
Duran, A.~G.; and Niepert, M. 2017.
\newblock Learning graph representations with embedding propagation.
\newblock In \emph{NIPS}, 5119--5130.

\bibitem[{Easley and Kleinberg(2010)}]{easley2010networks}
Easley, D.; and Kleinberg, J. 2010.
\newblock \emph{Networks, crowds, and markets: Reasoning about a highly
  connected world}, volume~8.
\newblock Cambridge University Press.

\bibitem[{Gilmer et~al.(2017)Gilmer, Schoenholz, Riley, Vinyals, and
  Dahl}]{gilmer2017neural}
Gilmer, J.; Schoenholz, S.~S.; Riley, P.~F.; Vinyals, O.; and Dahl, G.~E. 2017.
\newblock Neural message passing for Quantum chemistry.
\newblock In \emph{ICML}, 1263--1272.

\bibitem[{Grover and Leskovec(2016)}]{grover2016node2vec}
Grover, A.; and Leskovec, J. 2016.
\newblock node2vec: Scalable feature learning for networks.
\newblock In \emph{KDD}, 855--864. ACM.

\bibitem[{Hamilton, Ying, and Leskovec(2017)}]{hamilton2017inductive}
Hamilton, W.; Ying, Z.; and Leskovec, J. 2017.
\newblock Inductive representation learning on large graphs.
\newblock In \emph{NIPS}, 1025--1035.

\bibitem[{Heider(1946)}]{heider1946attitudes}
Heider, F. 1946.
\newblock Attitudes and cognitive organization.
\newblock \emph{The Journal of psychology} 21(1): 107--112.

\bibitem[{Huang and Luo(2018)}]{huang2018computing}
Huang, J.; and Luo, T. 2018.
\newblock Computing Len for Exploring the Historical People's Social Network.
\newblock In \emph{FiCloudW}, 95--101. IEEE.

\bibitem[{Huang et~al.(2019)Huang, Shen, Hou, and Cheng}]{huang2019signed}
Huang, J.; Shen, H.; Hou, L.; and Cheng, X. 2019.
\newblock Signed graph attention networks.
\newblock In \emph{ICANN}, 566--577. Springer.

\bibitem[{Islam, Prakash, and Ramakrishnan(2018)}]{islam2018signet}
Islam, M.~R.; Prakash, B.~A.; and Ramakrishnan, N. 2018.
\newblock Signet: Scalable embeddings for signed networks.
\newblock In \emph{PAKDD}, 157--169. Springer.

\bibitem[{Kim et~al.(2018)Kim, Park, Lee, and Kang}]{kim2018side}
Kim, J.; Park, H.; Lee, J.-E.; and Kang, U. 2018.
\newblock Side: representation learning in signed directed networks.
\newblock In \emph{WWW}, 509--518.

\bibitem[{Kipf and Welling(2016)}]{kipf2016variational}
Kipf, T.~N.; and Welling, M. 2016.
\newblock Variational graph auto-encoders.
\newblock In \emph{NIPS Workshop on Bayesian Deep Learning}.

\bibitem[{Kipf and Welling(2019)}]{kipf2016semi}
Kipf, T.~N.; and Welling, M. 2019.
\newblock Semi-supervised classification with graph convolutional networks.
\newblock In \emph{ICLR}.

\bibitem[{Kumar et~al.(2018)Kumar, Hamilton, Leskovec, and
  Jurafsky}]{kumar2018community}
Kumar, S.; Hamilton, W.~L.; Leskovec, J.; and Jurafsky, D. 2018.
\newblock Community interaction and conflict on the web.
\newblock In \emph{WWW}, 933--943.

\bibitem[{Kumar et~al.(2016)Kumar, Spezzano, Subrahmanian, and
  Faloutsos}]{kumar2016edge}
Kumar, S.; Spezzano, F.; Subrahmanian, V.; and Faloutsos, C. 2016.
\newblock Edge weight prediction in weighted signed networks.
\newblock In \emph{ICDM}, 221--230. IEEE.

\bibitem[{Leskovec, Huttenlocher, and
  Kleinberg(2010{\natexlab{a}})}]{leskovec2010predicting}
Leskovec, J.; Huttenlocher, D.; and Kleinberg, J. 2010{\natexlab{a}}.
\newblock Predicting positive and negative links in online social networks.
\newblock In \emph{WWW}, 641--650. ACM.

\bibitem[{Leskovec, Huttenlocher, and
  Kleinberg(2010{\natexlab{b}})}]{leskovec2010signed}
Leskovec, J.; Huttenlocher, D.; and Kleinberg, J. 2010{\natexlab{b}}.
\newblock Signed networks in social media.
\newblock In \emph{CHI}, 1361--1370. ACM.

\bibitem[{Li et~al.(2020)Li, Tian, Zhang, and Chang}]{li2020learning}
Li, Y.; Tian, Y.; Zhang, J.; and Chang, Y. 2020.
\newblock Learning Signed Network Embedding via Graph Attention.
\newblock In \emph{AAAI}, 4772--4779.

\bibitem[{Li et~al.(2018)Li, Yuan, Wu, and Lu}]{li2016spectral}
Li, Y.; Yuan, S.; Wu, X.; and Lu, A. 2018.
\newblock On spectral analysis of directed signed graphs.
\newblock \emph{International Journal of Data Science and Analytics} 6(2):
  147--162.

\bibitem[{Pan et~al.(2018)Pan, Hu, Long, Jiang, Yao, and
  Zhang}]{pan2018adversarially}
Pan, S.; Hu, R.; Long, G.; Jiang, J.; Yao, L.; and Zhang, C. 2018.
\newblock Adversarially Regularized Graph Autoencoder for Graph Embedding.
\newblock In \emph{IJCAI}.

\bibitem[{Pedregosa et~al.(2011)Pedregosa, Varoquaux, Gramfort, Michel,
  Thirion, Grisel, Blondel, Prettenhofer, Weiss, Dubourg, Vanderplas, Passos,
  Cournapeau, Brucher, Perrot, and Duchesnay}]{scikit-learn}
Pedregosa, F.; Varoquaux, G.; Gramfort, A.; Michel, V.; Thirion, B.; Grisel,
  O.; Blondel, M.; Prettenhofer, P.; Weiss, R.; Dubourg, V.; Vanderplas, J.;
  Passos, A.; Cournapeau, D.; Brucher, M.; Perrot, M.; and Duchesnay, E. 2011.
\newblock Scikit-learn: Machine Learning in {P}ython.
\newblock \emph{JMLR} 12: 2825--2830.

\bibitem[{Perozzi, Al-Rfou, and Skiena(2014)}]{perozzi2014deepwalk}
Perozzi, B.; Al-Rfou, R.; and Skiena, S. 2014.
\newblock Deepwalk: Online learning of social representations.
\newblock In \emph{KDD}, 701--710. ACM.

\bibitem[{Schlichtkrull et~al.(2018)Schlichtkrull, Kipf, Bloem, Van Den~Berg,
  Titov, and Welling}]{schlichtkrull2018modeling}
Schlichtkrull, M.; Kipf, T.~N.; Bloem, P.; Van Den~Berg, R.; Titov, I.; and
  Welling, M. 2018.
\newblock Modeling relational data with graph convolutional networks.
\newblock In \emph{European Semantic Web Conference}, 593--607. Springer.

\bibitem[{Tang, Aggarwal, and Liu(2016)}]{tang2016node}
Tang, J.; Aggarwal, C.; and Liu, H. 2016.
\newblock Node classification in signed social networks.
\newblock In \emph{SIAM}, 54--62. SIAM.

\bibitem[{Tang et~al.(2015)Tang, Qu, Wang, Zhang, Yan, and Mei}]{tang2015line}
Tang, J.; Qu, M.; Wang, M.; Zhang, M.; Yan, J.; and Mei, Q. 2015.
\newblock Line: Large-scale information network embedding.
\newblock In \emph{WWW}, 1067--1077.

\bibitem[{Traag and Bruggeman(2009)}]{traag2009community}
Traag, V.~A.; and Bruggeman, J. 2009.
\newblock Community detection in networks with positive and negative links.
\newblock \emph{Physical Review E} 80(3): 036115.

\bibitem[{Veli{\v{c}}kovi{\'{c}} et~al.(2018)Veli{\v{c}}kovi{\'{c}}, Cucurull,
  Casanova, Romero, Li{\`{o}}, and Bengio}]{velickovic2017graph}
Veli{\v{c}}kovi{\'{c}}, P.; Cucurull, G.; Casanova, A.; Romero, A.; Li{\`{o}},
  P.; and Bengio, Y. 2018.
\newblock Graph attention networks.
\newblock \emph{ICLR}
  \urlprefix\url{https://openreview.net/forum?id=rJXMpikCZ}.

\bibitem[{Wang, Cui, and Zhu(2016)}]{wang2016structural}
Wang, D.; Cui, P.; and Zhu, W. 2016.
\newblock Structural deep network embedding.
\newblock In \emph{KDD}, 1225--1234. ACM.

\bibitem[{Wang et~al.(2017)Wang, Tang, Aggarwal, Chang, and
  Liu}]{wang2017signed}
Wang, S.; Tang, J.; Aggarwal, C.; Chang, Y.; and Liu, H. 2017.
\newblock Signed network embedding in social media.
\newblock In \emph{SIAM}, 327--335. SIAM.

\bibitem[{Wasserman, Faust et~al.(1994)}]{wasserman1994social}
Wasserman, S.; Faust, K.; et~al. 1994.
\newblock \emph{Social network analysis: Methods and applications}, volume~8.
\newblock Cambridge University Press.

\bibitem[{Wu et~al.(2020)Wu, Pan, Chen, Long, Zhang, and
  Philip}]{wu2019comprehensive}
Wu, Z.; Pan, S.; Chen, F.; Long, G.; Zhang, C.; and Philip, S.~Y. 2020.
\newblock A comprehensive survey on graph neural networks.
\newblock \emph{IEEE Transactions on Neural Networks and Learning Systems} .

\bibitem[{Xu et~al.(2020)Xu, Huang, Hou, Shen, Gao, and Cheng}]{xu2020label}
Xu, B.; Huang, J.; Hou, L.; Shen, H.; Gao, J.; and Cheng, X. 2020.
\newblock Label-Consistency based Graph Neural Networks for Semi-supervised
  Node Classification.
\newblock In \emph{SIGIR}, 1897--1900.

\bibitem[{Xu et~al.(2019)Xu, Shen, Cao, Qiu, and Cheng}]{xu2019graph}
Xu, B.; Shen, H.; Cao, Q.; Qiu, Y.; and Cheng, X. 2019.
\newblock Graph wavelet neural network.
\newblock In \emph{ICLR}.

\bibitem[{Yuan, Wu, and Xiang(2017)}]{yuan2017sne}
Yuan, S.; Wu, X.; and Xiang, Y. 2017.
\newblock Sne: signed network embedding.
\newblock In \emph{PAKDD}, 183--195. Springer.

\end{thebibliography}

\section{Appendix}
 \subsection{Comparison of Different Model Architectures}
In general, our model is a layer-by-layer signed relation GNN model with social theories guided loss functions.
We list the difference between the proposed method with existing baselines in \tableref{tb:comparison}.

\begin{table}[!ht]
  \centering
  \caption{Summary of selected methods.}
  \label{tb:comparison}
  \begin{tabular}{cccc}
    \toprule
    Method & SDGNN & BESIDE & SiGAT \\ 
    \midrule
    GNN Aggregator & \cmark & \xmark & \cmark \\ \hline
    Layer by Layer & \cmark & \xmark & \xmark \\ \hline
    Sign Loss & \cmark & \xmark & \cmark \\ \hline
    Direction Loss & \cmark & \cmark & \xmark \\ \hline
    Triangle Loss & \cmark &\cmark & \xmark \\ 
    \bottomrule
  \end{tabular}
\end{table}

\subsection{Algorithm Detail}
When our model work on a large signed directed graph whose number of nodes is more than 100,000, we could not put the whole graph into our memory with $L$-layer GNN aggregators ($L \geq 1$).

To train our SGDNN model in such large networks, we use mini-batch stochastic gradient descent to update the parameters SDGNN.
It needs to recombine the neighborhoods of the nodes in the batch to achieves batch calculation.
In this paper, the batch size is 500.
The training procedure is summarized in \algref{alg:algorithm1}. 
\begin{algorithm}[!ht]
  \caption{SDGNN Algorithm}
  \label{alg:algorithm1}
  \begin{algorithmic}[1]
    \renewcommand{\algorithmicrequire}{\textbf{Input:}}
    \renewcommand{\algorithmicensure}{\textbf{Output:}}
    \REQUIRE {
      Signed Directed Graph $\G(\V,\E, s)$;\\
      Encoder Aggregators $Enc$;\\
      GNN Layer Number $L$;\\
      Epoch $T$;
    }
    
    \ENSURE{
      Node representation $Z$
    }
    \\
    \STATE{Prepare Original node embeddings $z^0_u, \forall u \in V$.}
    \STATE{Initialize the parameters of neural networks.}
    \FOR{$epoch=1,...,T$}
    \FOR{each mini-batch node-set $B$ from $\V$}
    \STATE{Get neighborhoods $\mathcal{N}(v),  \forall v \in B$}
    \FOR{$u \in B$}
    \FOR{$l=1...L$}
    \STATE{Get neighborhoods embeddings $Z^l_{\mathcal{N}(v)},  \forall v \in B$}
    \STATE{
      $z_u^{l+1} \leftarrow Enc^l(z_u^l, Z^l_{\mathcal{N}(u)})$
    }
    \ENDFOR
    \ENDFOR
    \STATE{$Z_{B} \leftarrow z_v^{L} , \forall v \in B$}
    \STATE{Compute loss $\sum\limits_{v \in B} \mathcal{L}_{loss}(v)$ with $Z_{B}$}
    \STATE{Back propagation, update parameters.}
    \ENDFOR
    \ENDFOR
    \RETURN {$Z = Enc^L(V)$}
  \end{algorithmic}
\end{algorithm}

From \algref{alg:algorithm1}, we can find that the number of computed neighbors increases exponentially, which is consistent with GraphSAGE\citepp{hamilton2017inductive}.
In this paper, we set $L = 2$, which can capture triangles structure information. 
\subsection{Experiment Detail}
As we said in Experimental Settings section, the dataset used in this paper is public and can be downloaded from the corresponding links.
Data preprocessing is also discussed in Experiment Setting section, including how to binarize the networks.
In wikirfa dataset, there exits duplicate links (\ie A $\rightarrow^{+}$ B and A $\rightarrow^{-}$ B ), we use the last record in the edge list.
That's why it is slightly different from \citea{chen2018bridge}.
Besides, we number the nodes to make string id into a number id, which is eaiser to be vectorized.
These numbers are completed in the data preprocessing and are same to all embedding methods.

\subsection{Complexity Analysis}
We give complexity analysis in this section.
Firstly, in the dataset given in Dataset section, no node attribute information is provided.
For graph neural network methods (\ie SGCN, SiGAT and SDGNN), there are usually two types of methods: using id embedding or using spectral analysis methods.
For our SDGNN, we use a learnable id embeddings for the initial node representation.
It's $O(|\V| * d)$, where $d$ is the embedding dimensions.
Then, when we have $L$-layer GNN aggregators, we have the $\textbf{W}$ whose complexity is $O(5 * d * d)$.
Besides, we have a two-layer $\mathrm{MLP}$ to encoder the memssage for neighborhoods $\mathcal{N}(v)$ and node representation $z_v$.
The complexity of this part is $O(2 * 5 * d * d)$.
In total, the complexity of SGNN is $O(|\V| * d + 10 * d * d)$.
For attention aggregators, it will include an additional attention vector, whose complexity is $O(2 * d)$.

\subsection{Various Values of $\lambda_1$ and $\lambda_2$}
We show the results of various value of $\lambda_1$ and $\lambda_2$ in \tableref{tb:various-lambdas}.
From \tableref{tb:various-lambdas}, we can find that $\lambda_1$ and $\lambda_2$ are not zero,
showing better performance. 
Since reconstructing the sign is the most fundamental task, we don't set it to zero.

\begin{table}[!ht]
  \centering
    \caption{Various Values of $\lambda_1$ and $\lambda_2$.}
  \label{tb:various-lambdas}
  \scalebox{0.8}{
  \begin{tabular}{c|ccccc}
    \toprule
    & Metric  & $\lambda_2 = 0$  & $\lambda_2 = 1$ & $\lambda_2 = 5$ & $\lambda_2 = 10$  \\ 
    \midrule
    \multirow{4}{*}{$\lambda_1 = 0$} & Micro-F1 & 0.9386 & 0.9415 & 0.9411 & 0.9413 \\
    & Binary-F1 & 0.9677 & 0.9690 & 0.9688 & 0.9689 \\
    & Macro-F1 & 0.6738 & 0.7210 & 0.7267 & 0.7272 \\
    & AUC &  0.8883 & 0.9030 & 0.8991 & 0.9008 \\ \hline
    \multirow{4}{*}{$\lambda_1 = 1$}  & Micro-F1 & 0.9438 & 0.9475 & 0.9452 & 0.9440 \\ 
    & Binary-F1 & 0.9702 & 0.9721 & 0.9709 & 0.9703 \\
    & Macro-F1 & 0.7414 & 0.7585 & 0.7484 & 0.7392 \\ 
    & AUC & 0.9082 & 0.9109 & 0.9070 & 0.9014 \\ \hline
    \multirow{4}{*}{$\lambda_1 = 5$}  & Micro-F1 & 0.9456 & 0.9440 & 0.9461 & 0.9436 \\
    & Binary-F1 & 0.9711 & 0.9702 & 0.9714 & 0.9700 \\ 
    & Macro-F1 & 0.7553 & 0.7495 & 0.7531 & 0.7476 \\
    & AUC & 0.9139 & 0.9152 & 0.9123 & 0.9088 \\ \hline
    \multirow{4}{*}{$\lambda_1 = 10$} & Micro-F1 & 0.9471 & 0.9465 & 0.9473 & 0.9452 \\
    & Binary-F1 & 0.9719 & 0.9715 & 0.9720 & 0.9709 \\
    & Macro-F1 & 0.7598 & 0.7598 & 0.7627 & 0.7526 \\
    & AUC & 0.9154 & 0.9162 & 0.9118 & 0.9127 \\ 
    \bottomrule
  \end{tabular}
  }

\end{table}

\end{document}